# Bilinear pooling and metric learning network for early Alzheimer's disease identification with FDG-PET images

Wenju Cui[1,2, (a], Caiying Yan[4, (a], Zhuangzhi Yan[1], Yunsong Peng[2,3], Yilin Leng[1,2], Chenlu Liu[4], Shuangqing Chen[4*], Xi Jiang[5*]

[1]Institute of Biomedical Engineering, School of Communication and Information Engineering, Shanghai University, Shanghai 200444, China

[2]Medical Imaging Department, Suzhou Institute of Biomedical Engineering and Technology, Chinese Academy of Sciences, Suzhou 215163, China

[3]School of Biomedical Engineering (Suzhou), Division of Life Sciences and Medicine, University of Science and Technology of China, Hefei 230026, China

[4]Department of Radiology, the Affiliated Suzhou Hospital of Nanjing Medical University, Suzhou 215163, China

[5]School of Life Science and Technology, The University of Electronic Science and Technology of China, Chengdu 611731, China

* Correspondence:
Shuangqing Chen, sznaonao@163.com;
Xi Jiang, xijiang@uestc.edu.cn

## Abstract

18F-fluorodeoxyglucose (FDG)-positron emission tomography (PET) reveals altered brain metabolism in individuals with mild cognitive impairment (MCI) and Alzheimer's disease (AD). Some biomarkers derived from FDG-PET by computer-aided-diagnosis (CAD) technologies have been proved that they can accurately diagnosis normal control (NC), MCI, and AD. However, the studies of identification of early MCI (EMCI) and late MCI (LMCI) with FDG-PET images are still insufficient. Compared with studies based on fMRI and DTI images, the researches of the inter-region representation features in FDG-PET images are insufficient. Moreover, considering the variability in different individuals, some hard samples which are very similar with both two classes limit the classification performance. To tackle these problems, in this paper, we propose a novel bilinear pooling and metric learning network (BMNet), which can extract the inter-region representation features and distinguish hard samples by constructing embedding space. To validate the proposed method, we collect 998 FDG-PET images from Alzheimer's disease neuroimaging initiative (ADNI) including 263 normal control (NC) patients, 290 EMCI patients, 147 LMCI patients, and 198 AD patients. Following the common preprocessing steps, 90 features are extracted from each FDG-PET image according to the automatic anatomical landmark (AAL) template and then sent into the proposed network. Extensive 5-fold cross-validation experiments are performed for multiple two-class classifications. Experiments show that most metrics are improved after adding the bilinear pooling module and metric losses to the Baseline model respectively. Specifically, in the classification task between EMCI and LMCI, the specificity improves 6.38% after adding the triple metric loss, and the negative predictive value (NPV) improves 3.45% after using the bilinear pooling module. In addition, the accuracy of classification between EMCI and LMCI achieves 79.64% using imbalanced FDG-PET images, which illustrates that the proposed method yields a state-of-the-art result of the classification accuracy between EMCI and LMCI based on PET images.

Keywords: early Alzheimer's disease; mild cognitive impairment; FDG-PET images; bilinear pooling; inter-region representation; metric learning; embedding space;

[a These authors contributed equally to this work.

# 1. INTRODUCTION

Alzheimer's disease (AD), a brain degenerative disorder, is harming the health of thousands of old people now, and its rate of prevalence is expected to increase rapidly in the coming decades [1-3]. Mild cognitive impairment (MCI) is considered to be a preclinical precursor of AD, but it is difficult to predict whether it will convert to AD or not [4-6]. Considering the unpredictable process of MCI, it is crucial to develop relevant methods for diagnosing the early MCI and AD.

18F-fluorodeoxyglucose (FDG)-positron emission tomography (PET) can reveal altered brain metabolism in individuals with MCI and AD [7-9]. Furthermore, various recent studies have proved that biomarkers derived from FDG-PET by computer-aided-diagnosis (CAD) technologies of machine learning and deep learning can accurately diagnose NC, MCI, and AD [10-12]. Liu et.al [13] proposed a new classification framework for AD diagnosis with 3D PET images. They decomposed 3D images into 2D slices to learn the intra-slice and inter-slice features and achieved a promising classification performance of AUC of 83.9% for MCI vs. NC classification. Zhou et.al [14] developed a new deep belief network model for AD diagnosis based on sparse-response theory, which identified a better classification result than that of other models. To solve the multimodal data missing problem, Dong et.al [15] proposed a high-order Laplacian regularized low-rank representation method for the classification tasks of NC, MCI, and AD.

Many studies have achieved good performance on the classification of NC, MCI, and AD based on FDG-PET images. However, when it comes to the more refined classification of early MCI (EMCI) and late MCI (LMCI), the studies with FDG-PET images are insufficient, especially compared with the relevant researches based on functional magnetic resonance imaging (fMRI) and diffusion tensor imaging (DTI) images. Hao et.al [16] proposed a novel multi-modal neuroimaging feature selection method with consistent metric constraint (MFCC) and obtained an accuracy (ACC) of 73.87% for the classification between EMCI and LMCI based on MRI and FDG-PET but only 64.69% when just using FDG-PET. Shibani et.al [17] proposed a multilayer neural network involving probabilistic principal component analysis for binary classification and only achieved an F1 score of 0.6844. Nozadi et.al [18] used learned features from semantically labeled PET images to perform group classification and got an ACC of 72.5%. Forouzannezhada et.al [19-20] applied a novel deep neural network and a random forest model respectively, and both models got a moderate ACC. Fang et.al [21] introduced a supervised Gaussian discriminative component analysis (GDCA) algorithm for the effective classification of early Alzheimer's disease with MRI and PET. Yang et.al [22] applied the Convolutional Architecture for Fast Feature Embedding (CAFFE) as the framework of the deep learning platform for early Alzheimer's disease diagnosis. Fang et.al [21] and Yang et.al [22] both gained some breakthrough of the classification results based on FDG-PET images but there was still a gap compared with these based on fMRI and DTI.

By comparison, based on fMRI and DTI images, Lei et.al [23] got an ACC of 78.05% for the classification between EMCI and LMCI via proposing a new joint multi-task learning method by combining low-rank self-calibrated functional and structural brain networks. Furthermore, Song et.al [24] constructed a new graph convolution network (GCN) and got an ACC of 79.26% based on fMRI and 82.92% based on DTI for the same classification task.

To sum up, there is indeed a gap in the classification between EMCI and LMCI based on FDG-PET images compared with fMRI and DTI images. One of the reasons might be that existing classification methods based on FDG-PET have not fully explored the inter-region representation among different brain regions. For example, based on fMRI, there are many methods like Pearson's correlation and sparse representation for functional brain network (FBN) estimation [25]. However, several studies have proved that brain metabolism connectivity has value in the diagnosis of early AD [26-28], but there are few studies to improve classification performance using the inter-region representation features based on FDG-PET. In addition, another reason might be that the number of PET images is generally much more than that of fMRI images in most researches. The bigger dataset might increase the variety of individuals and the probability of special samples which are hard to distinguish, thus causing complexity of the problem for classification tasks.

Considering these two limitations, we propose a novel bilinear pooling and metric learning network (BMNet) for early Alzheimer's disease identification with FDG-PET images, especially for the classification task between EMCI and LMCI. Our main contributions are as follows: **(1) We propose a shallow convolutional neural network model to achieve the classification; (2) We introduce a bilinear pooling module into the model for exploring the inter-region representation features in the whole brain; (3) We introduce the deep metric learning to help model learn the hard samples in the embedding feature space; (4) We conduct our method on the dataset collected from the publicly released ADNI database and obtain a state-of-the-art result of the classification between EMCI and LMCI based on PET images.**

The rest of this paper is organized as follows. In section 2, we present details of the materials and the proposed methods. Section 3 presents the results of the experiments on the public ADNI database. Finally, we provide the discussions and conclusion of this paper in section 4.

## 2. Materials and Methods

### 2.1 Image acquisition and preprocessing

**Table 1.** Demographic characteristics of the subjects in the ADNI database. The values are presented as mean ± standard deviation. MMSE represents Mini-Mental State Examination.

| Subjects | NC | EMCI | LMCI | AD |
| --- | --- | --- | --- | --- |
| Number | 263 | 290 | 147 | 198 |
| Gender (M/F) | 130/133 | 160/130 | 80/67 | 119/79 |
| Age | 75.49±6.47 | 71.40±7.33 | 72.16±7.55 | 75.05±7.60 |
| MMSE | 29.06±1.13 | 28.32±1.57 | 27.62±1.84 | 23.20±2.17 |

In this work, we use the data in the publicly released Alzheimer's Disease Neuroimaging Initiative (ADNI) database [29]. We collect a cohort of subjects with FDG-PET images from the ADNI databases. The ADNI cohort includes FDG-PET images from 898 subjects, including 263 NC, 290 EMCI, 147 LMCI, and 198 AD participants. Table 1 lists the demographic characteristics of subjects.

We choose FDG-PET images which are in a state of rest with 30-35 min with $185 \pm 18.5$ MBq FDG, and details of acquisition can be obtained from the study protocols in the ADNI database. Firstly, we normalize the images based on the template of the Montreal Neurological Institute (MNI). Then, we perform the smoothing with a Gaussian filter of 8 mm fullwidth at half-maximum (FWHM) [9]. Finally, in order to verify the effectiveness of the proposed method, we do the main experiments using two different brain atlas. Based on the automated anatomical labeling (AAL) [30] atlas, we extract features of 90 regions of interest (ROIs) from FDG-PET images with intensity normalized averagely. Similarly, based on the Schaefer 2018 atlas [24.5], we extract features of 400 regions. We perform all preprocessing steps by Statistical Parametric Mapping software (SPM12) [32] and MATLAB 2020a [33].

## 2.2 Methods

### 2.2.1 Overview of the proposed network

Figure 1 illustrates the method framework of this study. The left box is the preprocessing step of FDG-PET images, in which the left image is the raw PET image of the brain, and the right one is the AAL template. Then 90 features extracted based on the AAL template are input into the subsequent model. The model consists of two convolution layers, a bilinear pooling layer, and two fully connected layers.

After extracting the first-order features through two convolution layers, the bilinear pooling module is used to further extract the inter-region features between brain regions. Finally, the metric learning loss is added to the classification loss to strengthen the ability to learn hard samples of the proposed model.

### 2.2.2 Baseline model

We construct a shallow neural network as the Baseline model, including two convolution blocks and three fully connected layers. Each convolution block includes a convolution layer, a batch normalization layer, and a Rectified Linear Unit (ReLU) activation layer.

Given a set of nodes (regions) R={r1,r2,...,r3}, and the features of each region is denoted as $X_i$. Each convolution block is defined as:

$$Y_i = \sigma(BN(f(X_i)))  \quad (1)$$

Where the $f$ represents the convolution process, $BN$ represents the batch normalization process, $\sigma$ represents the activation process.

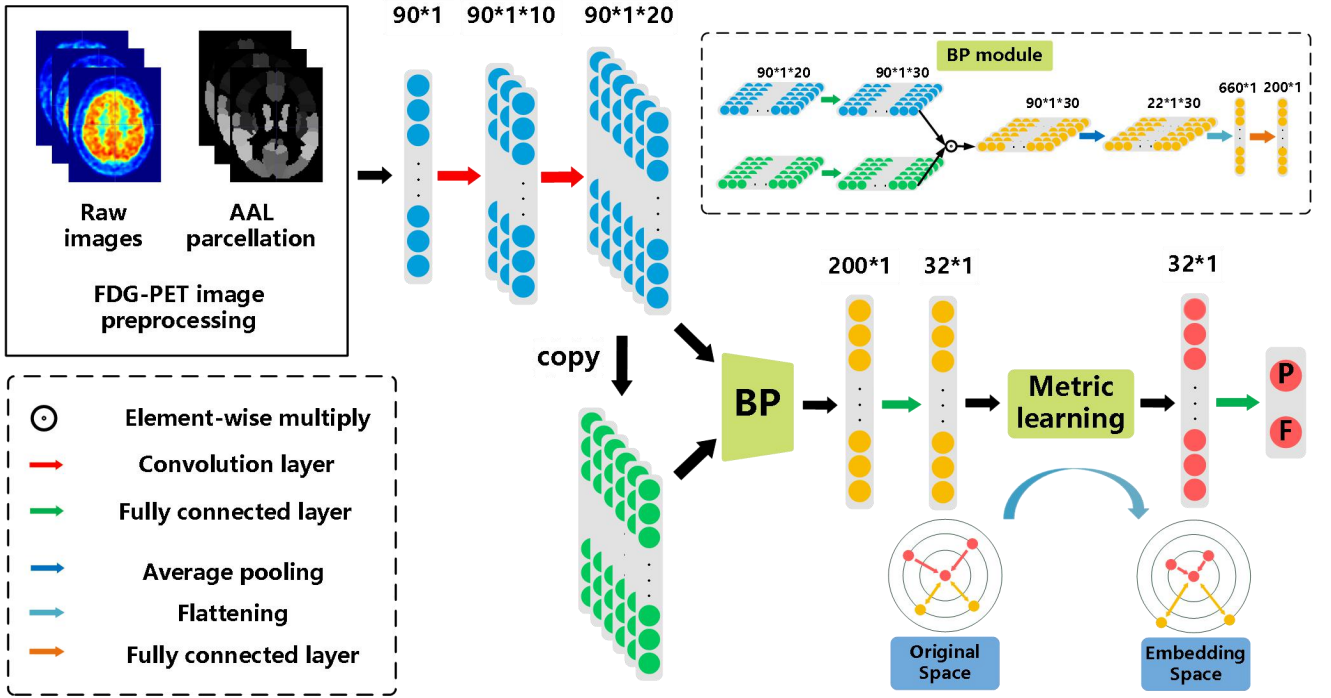

**Figure 1.** The architecture of the proposed bilinear pooling and metric learning network (BMNet) for MCI diagnosis using PET images. There are four modules in our framework (i.e., images preprocessing module, convolutional feature-extraction module, bilinear pooling module, and the metric loss module).

### 2.2.3 Generating inter-region representation via bilinear pooling module

In this section, we propose to use a bilinear pooling module to further generate second-order features which may represent inter-region features among whole brain regions [34]. Bilinear pooling is an effective feature fusion method, which has been widely used in various computer vision and machine learning tasks [35]. Bilinear pooling captures the relationship between features by modeling the high-order statistical information of features and then generates an expressive global representation [35]. And in the research of DTI and fMRI, this method is also used to extract connectivity-based features between brain regions [36]. In theory, by using these features, the inter-region representation among the whole brain regions in FDG-PET images could be exploited to some extent, as the functional brain network of fMRI.

In this work, we use bilinear pooling to capture the second-order features to generate inter-region features from FDG-PET images. The bilinear features are calculated as follows:

$$B_{ij} = f_1(Y_{ij}) \cdot f_2(Y_{ij}) \quad (2)$$

where $B_{ij}$ represents the bilinear features, and $Y_{ij}$ represents the inputs features, $f_1$ and $f_2$ are two fully connected layers.

After that, we use an average pooling layer to diminish the feature dimension and flatten the feature map to one-dimensional.

### 2.2.4 Distinguishing hard samples in embedding space by metric learning

In this section, we introduce the deep metric learning strategy into the classification of different stages of AD. Metric learning is widely utilized with deep neural networks in classification tasks, especially in problems affected by large intra-class sample changes [37,38]. Deep metric learning loss maps features to the embedded space, which is conducive to learning difficult samples and can effectively deal with the imbalance of data [38]. Inspired by these, we argue that deep metric learning might be suitable for our classification task. Thus, in this paper, we employ deep metric learning for the diagnosis of AD to help distinguish hard samples in the embedding space.

In deep neural networks, the loss function is a manifestation of metric learning, and there are a variety of different metric learning loss functions. In this paper, we employ two deep metric learning loss functions for automatic diagnosis of early AD, including contrastive loss and triplet loss, which are widely used in recent studies [38-40]. Contrastive loss employs a pair of positive and negative samples for each training iteration. The contrastive loss function is measured by the Euclidian distance between two vectors in embedding space. The contrastive loss function is given as [41]:

$$L_c(b_{1,i}, b_{2,i}) = \sum_{i=1}^{N} [y_i d_{1,2}^2 + (1 - y_i)\{max(0, m - d_{1,2})\}^2] \quad (3)$$

$$d_{1,2} = ||f_{1,i} - f_{2,i}||_2^2 \quad (4)$$

where $y_i = 0$ for two positive vectors and $y_i = 1$ for negative pairs, $b_{1,i}$, $b_{2,i}$ is the training input from two classes, $f_{1,i}$, $f_{2,i}$ represents the embedding vector of each training input generated by the network, $N$ is the number of input samples, and $m$ is the margin, usually set to 1.0.

When the input is a positive sample pair, $d_{1,2}$ decreases gradually, and the same kind of samples will continue to form clusters in the feature space. On the contrary, when the network inputs a negative sample pair, $d_{1,2}$ will gradually rise until it reaches the set $m$. By minimizing the loss functions, the distance between positive sample pairs can be gradually reduced and the distance between negative sample pairs can be gradually increased, to meet the needs of the classification task.

Triplet loss is a widely used measure of metric learning loss, which is the basis of a large number of metric learning methods. Unlike contrastive loss, triplet loss requires three input samples including two positive samples and a negative sample. The three samples are named as fixed sample (anchor) $b^a$, positive sample (positive) $b^p$ and negative sample (negative) $b^n$ respectively. $b^a$ and $b^p$ form positive sample pairs, and $b^a$ and $b^n$ form negative sample pairs.

This triplet loss function simultaneously penalizes a short distance $d_{a,n}$ between an anchor and a negative sample and a long distance $d_{a,p}$ between an anchor and a positive sample, and is defined as [42]:

$$L_{triplet}(b_i^a, b_i^p, b_i^n) = \sum_{i=1}^{N} max(0, m + d_{a,p} - d_{a,n}) \quad (5)$$

where $b_i^a$, $b_i^p$, $b_i^n$ is the input from two training groups, $N$ represents the number of samples, and $m$ is the margin, usually set to 1.0.

$$d_{a,p} = ||f_i^a - f_i^p||_2^2 \quad (6)$$

$$d_{a,n} = ||f_i^a - f_i^n||_2^2 \quad (7)$$

$f_i^a$, $f_i^p$, $f_i^n$ represents the vector of training input in embedding space.

As shown in Figure 1, triples can shorten the distance between positive sample pairs, while pushing away the distance between negative sample pairs. Finally, samples with the same class form feature clusters and embedding space to improve the performance of the classification tasks.

*2.2.5 Loss functions*

In addition, we use cross-entropy loss $L_C$ for the classification task. Therefore, the final loss function includes a joint loss function $L_{total}$ that contains metric loss $L_M$ for the embedding space and cross-entropy loss for the classification task.

$$L_{total} = \lambda L_M + L_C \quad (8)$$

$$L_C = \frac{1}{N}\sum_i^N -[y_i log(p_i) + (1-y_i)log(1-p_i)] \quad (9)$$

Where $y_i$ represents the label of the sample $i$, where $p_i$ represents the probability that the sample $i$ is projected to be a positive class, $\lambda$ represents the coefficient which we define as 0.05 by experience.

*2.2.6 Performance evaluation*

We adopt six commonly used evaluation metrics to evaluate the performance of the models objectively, including accuracy (ACC), sensitivity (SEN), specificity (SPE), positive predictive value (PPV), negative predictive value (PPV), F1 score (F1), area under the receiver operating characteristic curve (AUC).

*2.2.7 Implementation Details*

We implement the proposed network based on the public platform PyTorch 1.8 and Intel Core i5-9400 CPU with 16GB memory. Besides, we adapt stochastic gradient descent (SGD) to optimize the model, in which momentum and weight decay are set to 0.9 and 0.001 respectively.

*2.2.8 Validation strategies and Statistic analysis methods*

To evaluate the effectiveness of the proposed model, we conduct a five-fold cross-validation strategy in all ablation and comparative experiments based on the AAL atlas. For each experiment, we divide data into five groups, and each group maintains the same proportion of two classes. In each fold experiment, four groups are used as train groups and another group is used as the test group. The detailed classification results on the ADNI database are summarized in Section 3.1.

In addition, we apply independent testing set strategy in the experiments based on Schaefer 2018 atlas. We divide the collected dataset from the ADNI database into a training set (80%), validation set (10%), and testing set (10%). The corresponding detailed classification results are summarized in Sections 3.2.

Similarly, to evaluate the effectiveness of the proposed model, we use two methods to validate the statistical significance including the t-test and DeLong test. In the experiments on the AAL atlas, we use the t-test. In the experiments on Schaefer 2018 atlas, we use the DeLong test.

## 3. RESULTS

### 3.1 Ablation Experiments

To verify the effect of the bilinear pooling module and the metric learning loss on the performance of the proposed model, we remove the bilinear pooling module and the metric learning mechanism loss from the proposed BMNet, respectively. In the first experiment (i.e., our method without a bilinear pooling module), we directly use a fully connected layer to replace the bilinear pooling module. In the second experiment (i.e., our method without metric learning losses), we just use the cross-entropy loss function. The details are as follows and the results are shown in Tables 2-5 and Figures 2-3.

**The ablation experiments of the BP module**

Firstly, we conduct the experiments based on the Baseline model. Then we conduct the experiments of adding a bilinear pooling (BP) module to the Baseline model. According to the results, after the BP module is added, the four groups of classification experimental results have been improved to a certain extent. Specifically, in classification experiments between EMCI and LMCI, ACC increases by 2.74%, and AUC increases by 2.97%. In classification experiments between NC and AD, the results are the best, where ACC increases by 3.69% and AUC increases by 2.12%. In addition, we also conduct experiments in the classification between NC and LMCI, LMCI, and AD. The results illustrate that the BP module has a good generalization ability in the different classification tasks.

Furthermore, we conduct comparative experiments to verify the effectiveness of the BP module based on metric learning loss. For example, in the classification experiments of EMCI and LMCI, after adding the BP model to the triplet loss (Tri-loss), ACC increases by 2.29%, and AUC increases by 1.74%.

**The ablation experiments of metric learning losses**

In this sub-section, we perform comparative experiments in terms of metric learning losses, including the triplet loss (Tri-loss) and the contrastive loss (Con-loss). We use two kinds of metric learning losses respectively, and the results illustrate that the two metric learning losses are both effective in different experiments. Specifically, in the classification experiments between EMCI and LMCI, ACC increases by 2.07% after adding the contrastive loss, which is a little higher than that of triplet loss. Similarly, in the classification experiments between NC and AD, ACC increases by 3.89%. In the classification experiments between NC and LMCI, the results of triplet loss improve more than these of contrastive loss, and ACC reaches 0.8049. On the contrary, in the classification experiment between LMCI and AD, the results of contrastive loss are better, where ACC reaches 0.8177 and AUC reaches 0.8297.

Finally, we use the t-test to measure the statistical significance comparing AUCs and the results are shown as p-value in Tables 2-5. We can see that most results of the two final models (Con-loss+BP and Tri-loss+BP) are statistically significant. In addition, we can also see that most F1 scores of the two final models are higher than these of other models in Figures 2-3.

**Table 2.** Results of the ablation studies of BP module and metric learning losses for EMCI vs. LMCI classification.

| Method | ACC | PPV | NPV | SEN | SPE | AUC | F1 | p |
|---|---|---|---|---|---|---|---|---|
| Baseline | 0.7574 | 0.8379 | 0.5984 | 0.8079 | 0.6492 | 0.7332 | 0.8226 | - |
| Baseline+BP | 0.7848 | 0.8621 | 0.6329 | 0.8224 | 0.7029 | 0.7629 | 0.8417 | 0.068 |
| Tri-loss | 0.7735 | 0.8793 | 0.5634 | 0.8049 | 0.713 | 0.7415 | 0.8405 | 0.342 |
| Tri-loss+BP | **0.7964** | **0.8931** | 0.6055 | 0.8184 | **0.7429** | 0.7589 | **0.8541** | **0.013*** |
| Con-loss | 0.7781 | 0.8655 | 0.5917 | 0.8088 | 0.7053 | 0.7387 | 0.8394 | 0.342 |
| Con-loss+BP | 0.7940 | 0.869 | **0.6467** | **0.8322** | 0.7243 | **0.7707** | 0.8501 | 0.079 |

**Table 3.** Results of the ablation studies of BP module and metric learning losses for NC VS. AD classification.

| Method | ACC | PPV | NPV | SEN | SPE | AUC | F1 | p |
|---|---|---|---|---|---|---|---|---|
| Baseline | 0.8525 | 0.9201 | 0.7633 | 0.8391 | 0.8799 | 0.9074 | 0.8777 | - |
| Baseline+BP | 0.8894 | 0.9352 | 0.8279 | 0.8792 | 0.9093 | 0.9286 | 0.9063 | 0.051 |
| Tri-loss | 0.8829 | 0.9354 | 0.8133 | 0.8692 | 0.9047 | 0.9279 | 0.9035 | 0.059 |
| Tri-loss+BP | **0.8980** | 0.9353 | **0.8481** | 0.8928 | 0.9120 | 0.9281 | 0.9111 | **0.032*** |
| Con-loss | 0.8914 | 0.9353 | 0.8335 | 0.8802 | 0.9076 | 0.9281 | 0.9069 | 0.088 |
| Con-loss+BP | **0.8980** | **0.9390** | 0.8432 | 0.8891 | **0.9140** | **0.9334** | **0.9135** | **0.029*** |

**Table 4.** Results of the ablation studies of BP module and metric learning loss for NC VS. LMCI classification.

| Method | ACC | PPV | NPV | SEN | SPE | AUC | F1 | p |
|---|---|---|---|---|---|---|---|---|
| Baseline | 0.7681 | 0.7886 | 0.7345 | 0.8450 | 0.6681 | 0.7527 | 0.8158 | - |
| Baseline+BP | 0.8000 | 0.8628 | 0.6871 | 0.8325 | 0.7445 | 0.7871 | 0.8474 | **<0.001*** |
| Tri-loss | 0.8049 | 0.8933 | 0.6460 | 0.8191 | 0.7768 | 0.7702 | 0.8546 | **0.007*** |
| Tri-loss+BP | **0.8220** | **0.8936** | 0.6942 | 0.8418 | **0.7848** | 0.7985 | **0.8669** | **<0.001*** |
| Con-loss | 0.7903 | 0.8322 | 0.7136 | 0.8431 | 0.7128 | 0.7841 | 0.8376 | **0.016*** |
| Con-loss+BP | 0.8146 | 0.8464 | **0.7202** | **0.8496** | 0.7605 | **0.8096** | 0.8480 | **0.001*** |

**Table 5.** Results of the ablation studies of BP module and metric learning loss for LMCI vs. AD classification.

| Method | ACC | PPV | NPV | SEN | SPE | AUC | F1 | p |
|---|---|---|---|---|---|---|---|---|
| Baseline | 0.7769 | 0.7416 | 0.8030 | 0.7457 | 0.8086 | 0.7964 | 0.7437 | - |
| Baseline+BP | 0.8006 | 0.7292 | **0.8537** | 0.7862 | 0.8129 | 0.8108 | 0.7566 | 0.243 |
| Tr-iloss | 0.8060 | 0.7425 | 0.8531 | **0.7977** | 0.8223 | 0.8040 | 0.7691 | 0.307 |
| Tri-loss+BP | 0.8118 | **0.7890** | 0.8285 | 0.7753 | **0.8441** | 0.8167 | 0.7821 | **0.022*** |
| Con-loss | 0.7971 | 0.7487 | 0.8331 | 0.7753 | 0.8200 | 0.8018 | 0.7618 | 0.327 |
| Con-loss+BP | **0.8177** | 0.7754 | 0.8491 | 0.7964 | 0.8384 | **0.8297** | 0.7857 | **0.028*** |

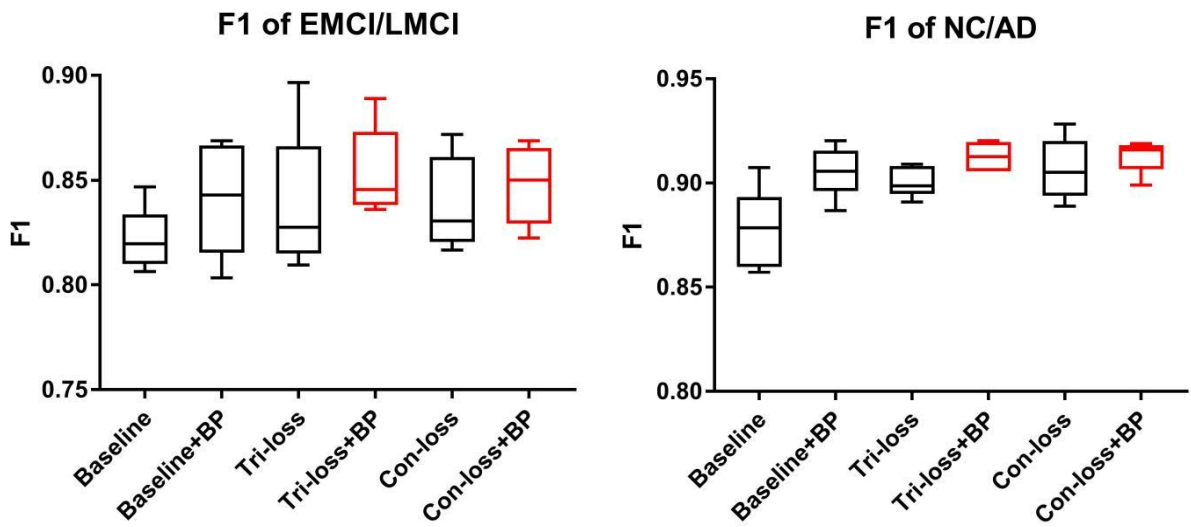

Figure 2. The F1 scores of experiments for EMCI vs. LMCI classification and the F1 scores of experiments for NC vs. LMCI classification.

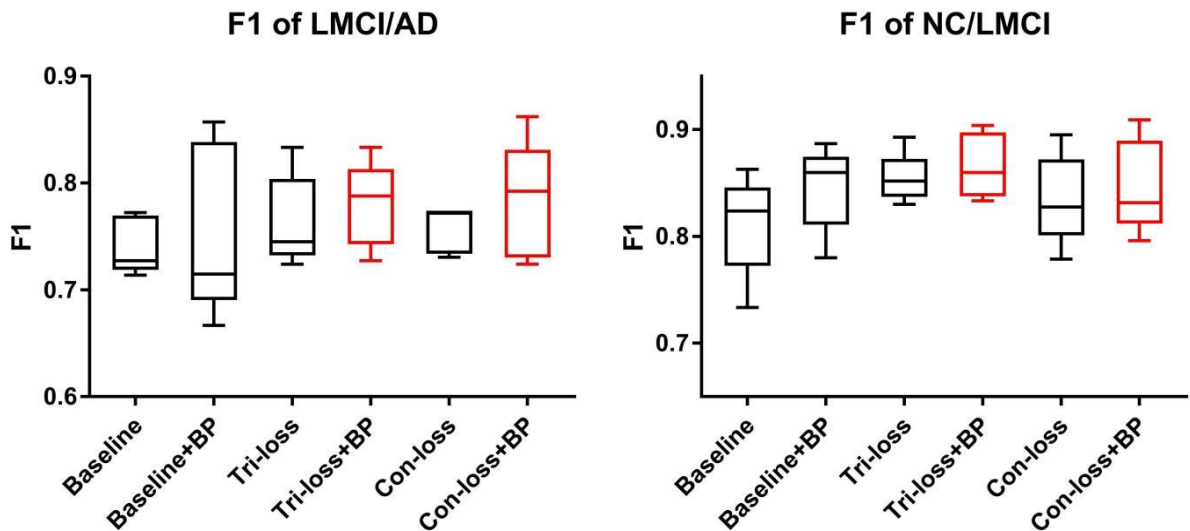

Figure 3. The F1 scores of experiments for LMCI vs. AD classification and the F1 scores of experiments for NC vs. AD classification.

## 3.2 Experiments on Differnet Atlases

In this section, we evaluate the performance of our method (Con-loss+BP) based on the Schaefer 2018 atlas. We conduct two groups of experiments for EMCI vs. LMCI classification and NC vs. AD classification and the results are shown in Table 10 and Figure 4. As stated earlier, we apply independent testing set strategy in these experiments and use the DeLong test to validate the statistical significance

The results illustrate that both BP module and contrastive loss are effective based on the Schaefer 2018 atlas. In the experiments for EMCI vs. LMCI classification, ACC increases by 2.27% after adding the contrastive loss, which is a little lower than that of the BP module. Similarly, in the classification experiments between NC and AD, ACC increases by 2.13%. Finally, combining the BP module and contrastive loss, the final model (Con-loss+BP) achieves much improvement in both two classification experiments. Specifically, in the classification experiments for EMCI and LMCI, ACC, SEN, SPE, F1 and AUC achieve 84.09%, 82.35%, 90%, 88.89% and 0.8529 with an improvement of 9.09%, 5.88%, 20%, 6.35% and 11.5% respectively, compared with Baseline model. In the NC vs. AD classification experiments, ACC, SEN, SPE, F1 and AUC increases by 6.38%, 4.1%, 9.47%, 5.72%, 7.78% and reach 89.36%, 89.29%, 89.47%, 90.91% and 0.9574.

**Table 10.** Results of the main studies based on the Schaefer 2018 atlas

| Class | Method | ACC | SEN | SPE | F1 | AUC | p |
|---|---|---|---|---|---|---|---|
| EMCI-LMCI | Baseline | 0.7500 | 0.7647 | 0.7000 | 0.8254 | 0.7379 | **0.0358*** |
|  | Con-loss | 0.7727 | 0.7714 | 0.7778 | 0.8438 | 0.7609 | 0.1090 |
|  | Baseline+BP | 0.7955 | 0.8125 | 0.7500 | 0.8525 | 0.7425 | 0.0990 |
|  | Con-loss+BP | **0.8409** | **0.8235** | **0.9000** | **0.8889** | **0.8529** | - |
| NC-AD | Baseline | 0.8298 | 0.8519 | 0.8000 | 0.8519 | 0.8796 | **0.0395*** |
|  | Con-loss | 0.8511 | 0.8571 | 0.8421 | 0.8727 | 0.9139 | 0.3212 |
|  | Baseline+BP | 0.8511 | 0.8333 | 0.8824 | 0.8772 | 0.9259 | 0.3548 |
|  | Con-loss+BP | **0.8936** | **0.8929** | **0.8947** | **0.9091** | **0.9574** | - |

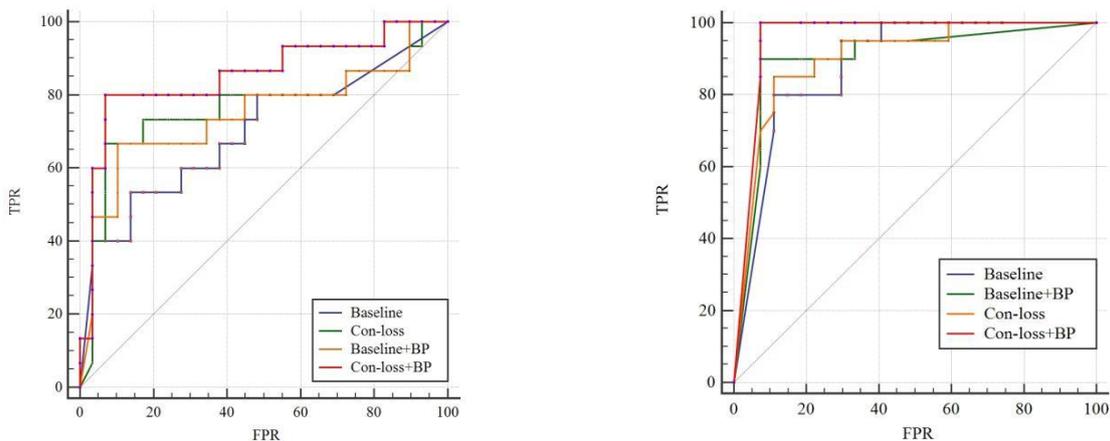

Figure 4. Receiver operating characteristic (ROC) curves of experiments for EMCI vs. LMCI classification and the ROC of experiments for NC vs. AD classification based on the Schaefer 2018 atlas. TPR, true positive rate; FPR, false-positive rate; AUC, area under the receiver operating characteristic curve. Please see the web version for the complete colorful picture.

## 3.3 Comparison With Other Methods

In this section, we compare the performance of our method (Tri-loss+BP) with that of several recent representative methods. In addition, we apply the least absolute shrinkage and selection operator (LASSO) feature selection method and support vector machine (SVM) method for the contrast experiments. From Table 6, we can find that our method gets the highest performance in classification experiments between EMCI and LMCI based on FDG-PET images. Specifically, the proposed method yields big improvement than the results of Singh et.al [17] and Nozadi et.al [18], although the dataset in our experiments is highly unbalanced. Based on a similar dataset, the proposed method still has better performance than the methods proposed by Forouzannezhad et.al [19-20]. In addition, compared with the method proposed by Hao et.al [16], our method achieves an overall huge improvement with 14.95% in ACC, 3.67% in SEN, 29.85% in SPE, and 12.89% in AUC, respectively. Compared to the results of the fusion of PET and MRI [16-21], our method also achieves an improvement in most metrics. Besides, our method gets a comparable performance compared to the methods based on fMRI and DTI adapted by Lei et. al [23] and by Song et. al [24], but the subjects in our research are much more than those they use.

**Table 6.** Comparison of the performance of different model algorithms in experiments for EMCI vs. LMCI classification with the related works.

| Method | Modality | DATA(EMCI/LMCI) | ACC | SEN | SPE | AUC | F1 |
|---|---|---|---|---|---|---|---|
| SVM | PET | 290/147 | 0.6620 | 0.7769 | 0.4653 | 0.6329 | - |
| Singh et.al [17] | PET | 178/158 | - | 0.6482 | - | - | 0.6844 |
| Nozadi et.al [18] | PET | 164/189 | 0.7250 | 0.7920 | 0.6990 | 0.790 | - |
| Forouzannezhad et.al [19] | PET | 296/193 | 0.6230 | 0.7820 | 0.4000 | - | - |
| Forouzannezhad et.al [20] | PET | 296/193 | 0.6280 | 0.6150 | 0.6430 | - | - |
| Yang et.al [22] | PET | - | 0.7219 | 0.7382 | 0.7305 | - | - |
| Hao et.al [16] | PET | 273/187 | 0.6469 | 0.7817 | 0.4444 | 0.6300 | - |
|  | PET+MRI |  | 0.7387 | 0.9055 | 0.4952 | 0.7000 | - |
| Fang et.al [21] | PET+MRI | 297/196 | 0.8333 | 0.8235 | 0.8966 | 0.8947 |  |
| Lei et.al [23] | fMRI | 44/38 | 0.7805 | 0.7368 | 0.8182 | 0.8571 | - |
|  | DTI |  | 0.5366 | 0.5789 | 0.5000 | 0.5260 | - |
| Song et.al [24] | fMRI | 44/38 | 0.7926 | 0.8421 | 0.7500 | 0.9067 | - |
|  | DTI |  | **0.8292** | **0.9473** | 0.7272 | **0.9414** | - |
| Our method (Tri-loss+BP) | PET | 290/147 | 0.7964 | 0.8184 | 0.7429 | 0.7589 | **0.8541** |
| Our method (Con-loss+BP on Schaefer atlas) | PET | 290/147 | **0.8409** | 0.8235 | **0.9000** | **0.8889** | 0.8529 |

Similarly, from Table 7, we can find that our method gets the highest performance of classification experiments between NC and AD based on PET images too. Specifically, compared with the method proposed by Hao et.al [16] based on PET images, our method achieves an overall huge improvement with 9.74% in ACC, 3.26% in SEN, 19.26% in SPE, and 7.81% in AUC, respectively. Besides, our method gets a comparable performance compared to the methods based on fMRI and DTI adapted by Lei et. al [23] and by Song et. al [24]. ACC, SEN, SPE, AUC of our method based on PET images improve 10.76%, 4.69%, 16.83%, and 2.82% than those of their method based on fMRI. While SEN and AUC are slightly lower, ACC and SPE based on PET images improve 7.1% and 19.11% than those based on DTI.

In addition, we conduct the classification experiments between NC and LMCI, LMCI and AD, and the results compared with other methods are shown in Table 8 and Table 9 respectively, to further validate the effectiveness of our method.

Table 7. Comparison of the performance of different model algorithms in experiments for NC vs. AD classification with the related works.

| Method | Modality | DATA(NC/AD) | ACC | SEN | SPE | AUC |
| --- | --- | --- | --- | --- | --- | --- |
| SVM | **PET** | 263/198 | 0.6213 | 0.8063 | 0.5547 | 0.8445 |
| Hao et. al [16] | **PET** | 211/160 | 0.8006 | 0.8602 | 0.7194 | 0.85 |
|  | MRI |  | 0.8663 | 0.9028 | 0.8181 | 0.93 |
| Lei et. al [23] | fMRI | 44/38 | 0.7805 | 0.7368 | 0.8182 | 0.8571 |
|  | DTI |  | 0.5366 | 0.5789 | 0.5000 | 0.5260 |
| Song et. al [24] | fMRI | 44/38 | 0.7926 | 0.8421 | 0.75 | 0.9067 |
|  | DTI |  | 0.8292 | **0.9473** | 0.7272 | 0.9414 |
| Our method (Tri-loss+BP) | **PET** | 263/198 | **0.898** | 0.8928 | **0.912** | 0.9281 |
| Our method (Con-loss+BP on Schaefer atlas) | **PET** | 263/198 | 0.8936 | 0.8929 | 0.8947 | **0.9574** |

Table 8. Comparison of the performance of different model algorithms in experiments for NC vs. LMCI classification with the related works.

| Method | Modality | DATA(NC/LMCI) | ACC | SEN | SPE | AUC |
| --- | --- | --- | --- | --- | --- | --- |
| SVM | **PET** | 263/147 | 0.6415 | 0.7446 | 0.5437 | 0.6724 |
| Hao et.al[16] | **PET** | 273/187 | 0.6677 | 0.7545 | 0.5594 | 0.68 |
|  | MRI |  | 0.712 | 0.7801 | 0.6332 | 0.76 |
| Lei et.al [23] | fMRI | 44/38 | 0.7805 | 0.7368 | **0.8182** | 0.8571 |
|  | DTI |  | 0.5366 | 0.5789 | 0.5000 | 0.5260 |
| Song et.al [24] | fMRI | 44/38 | 0.7926 | 0.8421 | 0.75 | 0.9067 |
|  | DTI |  | **0.8292** | **0.9473** | 0.7272 | **0.9414** |
| Our method (Tri-loss+BP) | **PET** | 263/147 | 0.822 | 0.8418 | 0.7848 | 0.7985 |

Table 9. Comparison of the performance of different model algorithms in experiments for LMCI vs. AD classification with the related works.

| Method | Modality | DATA(LMCI/AD) | ACC | SEN | SPE | AUC |
| --- | --- | --- | --- | --- | --- | --- |
| SVM | **PET** | 147/198 | 0.5841 | 0.7834 | 0.5044 | 0.6908 |
| Hao et.al[16] | **PET** | 273/187 | 0.6677 | 0.7545 | 0.5594 | 0.68 |
|  | MRI |  | 0.712 | 0.7801 | 0.6332 | 0.76 |
| Lei et. al [23] | fMRI | 44/38 | 0.7805 | 0.7368 | 0.8182 | 0.8571 |
|  | DTI |  | 0.5366 | 0.5789 | 0.5000 | 0.5260 |
| Song et. al [24] | fMRI | 44/38 | 0.7926 | 0.8421 | 0.75 | 0.9067 |
|  | DTI |  | **0.8292** | **0.9473** | 0.7272 | **0.9414** |
| Our method (Tri-loss+BP) | **PET** | 147/198 | 0.8118 | 0.7753 | **0.8441** | 0.8167 |

From those experiments above, we can see that our classification results between EMCI and LMCI have exceeded those of the existing methods overall based on FDG-PET images. In addition, our results are also comparable with those based on fMRI and DTI images.

# 4. Discussions and Conclusion

**Advances in current research**

In this paper, we propose a novel bilinear pooling and metric learning network (BMNet) for early Alzheimer's disease identification with FDG-PET images, especially for the classification task between EMCI and LMCI. Based on the brain regions, we propose a simple shallow neural network that is more effective than existing machine learning methods and more lightweight than existing deep learning methods. Furthermore, we focus on exploring the relationship between brain regions in FDG-PET images and apply advanced metric learning methods to target difficult samples. Experiments show that the model has a good performance in several classification tasks with FDG-PET images based on the ADNI database.

**Comparisons with previous research**

In general, there are two major advances between the proposed method and previous methods. Firstly, there are few studies to improve classification performance using the inter-region representation features based on FDG-PET. The capability of these methods is limited for this reason and cannot fully explore the inter-region representation among different brain regions. By comparison, the proposed BMNet introduces a bilinear pooling module into the model which can explore the inter-region representation features in the whole brain. Secondly, there are few methods to study hard samples to improve the classification results in the brain disorder analysis. By comparison, our method uses the metric learning method which has been proved useful for hard samples classification. We apply two metric learning losses to our model and they both get a good performance in the experiments.

**Possible application**

Considering the performance on FDG-PET images of the ADNI database, the proposed BMNet including the bilinear pooling module and the metric learning loss functions has the potential capability of diagnosis for other neurological diseases with other kinds of brain images. In the future, we will explore more applications of the proposed method.

**Limitations and future work**

While the proposed BMNet achieves good results for the diagnosis of early AD, there are still some limitations. Considering the characteristics of the convolution neural network, the models and results are hard to be interpreted and the relationship of the brain regions is difficult to be visualized. In addition, there is still some potential in exploiting methods that can extract inter-region representation features between brain regions based on FDG-PET images. In the future, we will try to design methods that could extract inter-region representation features more effectively.

**Conclusion**

In this study, we propose a novel neural network method for the diagnosis of early AD with FDG-PET. We firstly construct a shallow neural network as the Baseline model. Then we introduce a bilinear pooling module into the network to try to extract inter-region representation features among the whole brain. We also apply the deep metric learning losses into the final loss function to help distinguish hard samples in the embedding space. Finally, we conduct the BMNet on the ADNI database and the results show that our method yields comparable classification performance with several representative methods. Especially, we get a satisfying classification performance in the experiment between EMCI and LMCI, which is the state-of-the-art result with FDG-PET.

clustering[C]//Proceedings of the IEEE conference on computer vision and pattern recognition. 2015: 815-823.